# Element abundance patterns in stars indicate fission of nuclei heavier than uranium

Ian U. Roederer[1,*,†], Nicole Vassh[2], Erika M. Holmbeck[3], Matthew R. Mumpower[4,5], Rebecca Surman[6], John J. Cowan[7], Timothy C. Beers[6], Rana Ezzeddine[8], Anna Frebel[9,10], Terese T. Hansen[11], Vinicius M. Placco[12], Charli M. Sakari[13]

[1]Department of Astronomy, University of Michigan; Ann Arbor, MI 48109, USA.

[2]TRIUMF (Tri-University Meson Facility), Vancouver, BC V6T 2A3, Canada.

[3]Carnegie Observatories; Pasadena, CA 91101, USA.

[4]Theoretical Division, Los Alamos National Laboratory; Los Alamos, NM 87545, USA.

[5]Center for Theoretical Astrophysics, Los Alamos National Laboratory; Los Alamos, NM 87545, USA.

[6]Department of Physics and Astronomy, University of Notre Dame; Notre Dame, IN 46556, USA.

[7]Homer L. Dodge Department of Physics and Astronomy, University of Oklahoma; Norman, OK 73019, USA.

[8]Department of Astronomy, University of Florida, Bryant Space Science Center; Gainesville, FL 32611, USA.

[9]Department of Physics, Massachusetts Institute of Technology; Cambridge, MA 02139, USA.

[10]Kavli Institute for Astrophysics and Space Research, Massachusetts Institute of Technology; Cambridge, MA 02139, USA.

[11]Department of Astronomy, Stockholm University, AlbaNova University Center, SE-106 91 Stockholm, Sweden.

[12]National Optical-Infrared Astronomy Research Laboratory, National Science Foundation, Tucson, AZ 85719, USA.

[13]Department of Physics and Astronomy, San Francisco State University; San Francisco, CA 94132, USA.

*Corresponding author. Email: iuroederer@ncsu.edu

†Present address: Department of Physics, North Carolina State University; Raleigh, NC, USA.

**The heaviest chemical elements are naturally produced by the rapid neutron-capture process (r-process) during neutron star mergers or supernovae. The r-process production of elements heavier than uranium (transuranic nuclei) is poorly understood and inaccessible to experiments, so must be extrapolated using nucleosynthesis models. We examine element abundances in a sample of stars that are enhanced in r-process elements. The abundances of elements Ru, Rh, Pd, and Ag (atomic numbers $Z$ = 44 to 47, mass numbers $A$ = 99 to 110) correlate with those of heavier elements ($63 \leq Z \leq 78$, $A > 150$). There is no correlation for neighboring elements ($34 \leq Z \leq 42$ and $48 \leq Z \leq 62$). We interpret this as evidence that fission fragments of transuranic nuclei contribute to the abundances. Our results indicate that neutron-rich nuclei with mass numbers >260 are produced in r-process events.**

The heaviest chemical elements are synthesized through the rapid neutron capture process (r-process). Sites where the r-process occurs include mergers of neutron stars, which have been observed (*1-3*). The nuclei produced by the r-process in these events depend on the composition of material ejected by the merger (*4*) and the properties of the progenitor neutron stars, including their equation of state (*5*). Freshly produced lanthanide elements (atomic numbers $Z$ = 57 to 71) (*6, 7*) and Sr ($Z$ = 38) (*8*) have been observed in the ejecta of a neutron star merger event, but otherwise the detailed chemical composition has not been measured.

The detailed compositions of some ancient stars in the Milky Way have been determined from their spectra, which contain hundreds of absorption features of more than 40 r-process elements (*9*). The abundance patterns of lanthanide elements in these stars are nearly identical, indicating a possible universality of r-process events, producing the same abundance ratios. The composition of each star is dominated by the ejecta of individual r-process events (*10, 11*), such as neutron star mergers or rare types of supernova, which enriched the gas from which the stars formed (*12*).

**Stellar sample and abundance measurements**
We investigated the r-process using a sample of 42 stars in the Milky Way. We selected stars that were previously observed to have heavy elements known to be formed by the r-process, with no evidence of contamination from other processes [such as the slow neutron capture process (s-process)]. We adopted the element abundances of 31 heavy elements ($34 \leq Z \leq 90$) from data reported in 35 previous studies (*13*), then homogenized these abundances to a common scale using the atomic data reported in the original studies (*13*).

The ratio [Fe/H] is defined as [Fe/H] $\equiv \log_{10}(N_{Fe}) - \log_{10}(N_{Fe})_\odot$, where $N$ is the number density with subscript indicating the element, and $_\odot$ indicates the value for the Sun. We use Fe as a measure of the overall enrichment by elements heavier than helium (referred to as the metallicity). We use the ratio [Eu/Fe] $\equiv \log_{10}(N_{Eu}/N_{Fe}) - \log_{10}(N_{Eu}/N_{Fe})_\odot$ as a measure of the enhancement of r-process elements relative to the metallicity. Our sample spans the ranges $-3.57 \leq$ [Fe/H] $\leq -0.99$ and $-0.52 \leq$ [Eu/Fe] $\leq +1.69$ (Data S1).

**Correlations between elements**
Fig. 1 shows the heavy-element abundance patterns in the selected stars (individual elements are shown in Figs. S3 and S4). We find that stars with higher [Eu/Fe] ratios have abundances of some elements (including Ru, Rh, Pd, Ag, Gd, Tb, Dy, and Yb) that are slightly enhanced relative to stars with lower [Eu/Fe] ratios. This excess is not an expected consequence of r-process universality.

We calculated an empirical baseline abundance pattern (Tables S2 and S3) using the 13 stars (30% of the sample) with the lowest levels of r-process enhancement, [Eu/Fe] ≤ +0.3. The abundance excess, the difference between the individual abundance measurements in each star and the empirical baseline for that element, is shown for all stars in Figures 1B and D, and split into different groups of elements in Figure 2. Three sets of elements behave similarly in our sample: Se, Sr, Y, Zr, Nb, and Mo ($34 \leq Z \leq 42$); Cd, Sn, and Te ($48 \leq Z \leq 52$); and Ba, La, Ce, Pr, Nd, and Sm ($56 \leq Z \leq 62$), shown in Figs. 2A, 2C, and 2D, respectively. The abundance ratios of these elements exhibit no correlation with [Eu/Fe], and therefore no excess (*13*).

Two other sets of elements exhibit significant (at least 4.6σ) positive correlations with [Eu/Fe] (*13*). These two sets include the elements Ru, Rh, Pd, and Ag ($44 \leq Z \leq 47$); and Gd, Tb, Dy, Ho, Er, Tm, Yb, Hf, Os, and Pt ($64 \leq Z \leq 78$), shown in Fig. 2B and 2E respectively. These correlations indicate an extension of r-process universality, such that many of the heavier r-process elements have abundances linked to those of a smaller number of lighter elements.

**Interpretation as fission fragmentation**

Two deviations from r-process universality have previously been identified (see Supplementary Text). One consists of large differences (> 1.5 dex) in the overall abundances of light r-process elements ($34 \leq Z < 56$) relative to heavier ones ($Z \geq 56$) for some elements outside our range of interest ($44 \leq Z \leq 47$). The other deviation, known as the actinide boost, is characterized by small variations (< 0.7 dex) in the abundances of the actinide elements Th ($Z = 90$) and U ($Z = 92$) relative to other heavy r-process elements. Neither of these deviations from r-process universality can explain the correlations in Figure 2.

Our findings are also inconsistent with two-component models of r-process nucleosynthesis (*14, 15*), in which one component (referred to as the weak, or limited, r-process) dominates the lighter r-process elements and another (referred to as the main r-process) dominates the heavier r-process elements. Those two-component models segregate light and heavy element production, so cannot produce abundance correlations between [Eu/Fe] and some, but not all, light r-process elements.

We propose that these element groups ($44 \leq Z \leq 47$ and $63 \leq Z \leq 78$) were instead produced as fission fragments of transuranic ($Z > 92$) elements that were synthesized in the r-process but have since decayed. Nucleosynthesis models have predicted that transuranic elements can be produced in r-process events if the ejecta contain very neutron-rich material (*16-18*). In this scenario, the synthesis of heavy elements terminates at some maximum mass, above which transuranic nuclei undergo fission and increase the abundances of r-process elements at lower masses (*19*).

Fission fragments would reduce any variations in the initial abundances of r-process seed nuclei. Natural variations in conditions at nucleosynthesis sites would produce variations in the abundances, but fission fragment deposition could overcome those variations (*20*), leading to fixed relative abundances between the fission products (such as Ag and Eu). Fixed abundance ratios have previously been proposed as a potential signature of fission, based on theoretical calculations and a smaller sample of stars (*21*). Figure 3 shows theoretical models of neutron star merger dynamical ejecta (*13*), which produce similar log ε(Ag/Eu) ratios [defined as $\log_{10}(N_{Ag}/N_{Eu})$] when fission is included. Models that do not include a fission component near $A = 110$ (Fig. 3A) are sensitive to differences in initial conditions, such as neutron star mass, so predict variable log

ε(Ag/Eu) ratios in the ejecta. We refer specifically to fission products near $A = 110$, because all these models include some level of fission in their calculations.

Following a neutron star merger event, an accretion disk can temporarily form around the remnant object. Material ejected from this disk is expected to contain few or no nuclei that undergo fission (*22*), so additional ejecta from the disk could produce even larger variations between events. The neutron richness and relative amounts of ejecta from different parts of a neutron star merger event are not well constrained, so there is an unknown contribution of r-process synthesis in the disk to elements beyond Sr ($Z = 38$), such as Ag and Pd. We use previous dynamical ejecta simulations (*23*) to investigate the predictions of models that include a fission component for the mass-weighted log ε(Ag/Eu) ratios. We find they are approximately constant when fission is included (Fig 3B), which is consistent with the observed stellar ratios. We expect similar results for models of other potential r-process sites. For example, the predicted neutron richness of the ejecta of supernovae produced by magnetized and rapidly rotating stars (magneto-rotational supernovae) depends on the strength of the magnetic field (*24*). If fission occurs in any r-process site, it acts to reduce the variation in abundance ratios.

Figure 1 shows the baseline abundance pattern calculated above, which we assume contains minimal contributions from fission. To illustrate different fission contributions, Fig. 1 also shows abundance patterns with additional transuranic fission fragments enhanced by factors of 1, 2, and 4 times the baseline (*13*). These enhancement factors were chosen to span the range of abundance excesses we find for elements with $44 \leq Z \leq 47$ and $63 \leq Z \leq 78$ in the stars with the highest levels of [Eu/Fe]. We find that fission fragmentation could explain the elements with the largest observed excesses, including Pd, Ag, Gd, and Yb (*13*). We estimate that up to half of stars known to have $-3.0 \leq$ [Fe/H] $\leq -1.5$ could have contributions from fission fragments (*13*). Because the stars in our sample are old, this rate of fission-affected abundances is a property of the dominant r-process site in the early Universe.

We conclude that fission fragments of transuranic nuclei can explain the correlations we found between elements with $44 \leq Z \leq 47$ and $63 \leq Z \leq 78$. We investigated whether the observed behavior could be reproduced by other known nucleosynthesis processes, including the weak or main s-process, intermediate neutron-capture process, or weak r-process, but found it cannot (see Supplementary Text). Nor can it be explained by the actinide-boost phenomenon, which has previously been observed in some r-process-enhanced stars (Supplementary Text, Fig. S6).

Many of the radioactive, neutron-rich nuclei produced during the r-process are inaccessible to laboratory experiments, so theoretical models are necessary to estimate their properties by extrapolating from more stable isotopes. Models predict that many of these nuclei have asymmetric fission fragment distributions, with a lighter peak and a heavier peak (*13, 25*). We propose that the lighter fragments likely contribute to the Ru, Rh, Pd, and Ag abundances in our dataset. Neutron-rich nuclei with mass numbers $99 \leq A \leq 110$, produced directly through fission, are expected to experience β decay until they become the observed elements (*26*). If we assume that the heavier fragments contribute to the Eu and heavier elements ($A > 150$) in our sample, then nuclei with $A > 260$ (110 + 150) were produced in the r-process.

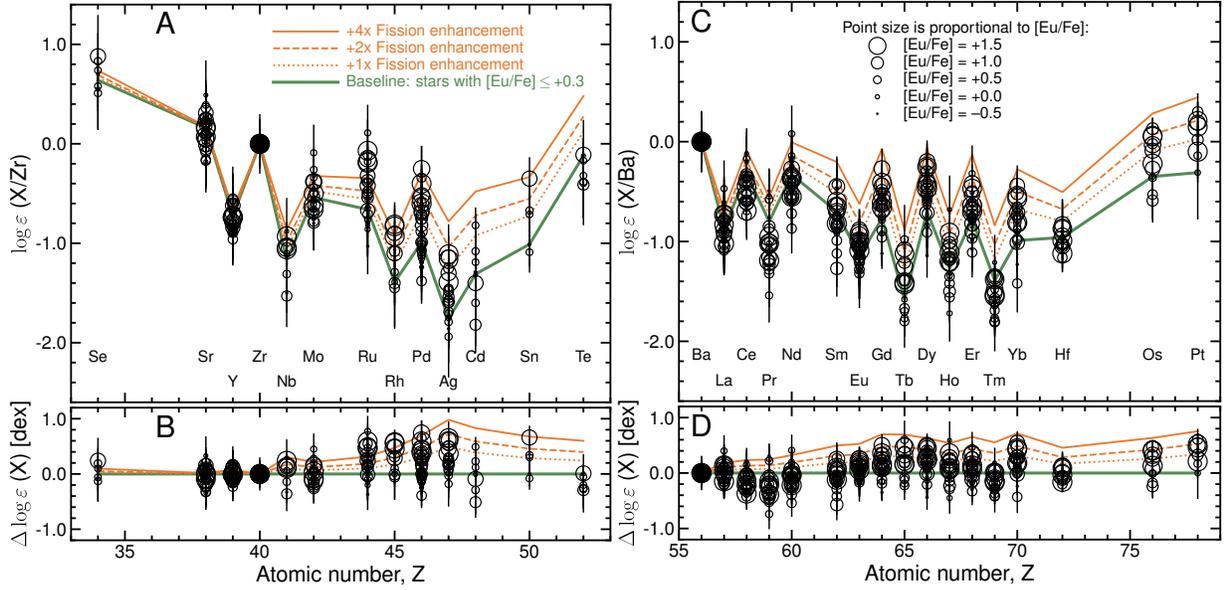

**Fig. 1: Observed abundance patterns compared to fission model predictions.** Logarithmic abundances (open circles) measured for 30 r-process elements in the 42 stars of our sample, plotted as a function of atomic number. The symbol sizes are proportional to [Eu/Fe] and error bars indicate 1σ uncertainties. The green line is the empirical baseline pattern we defined as the mean abundance ratios for the subset of 13 stars with [Eu/Fe] ≤ +0.3. Light and dark green shading indicates ± 1 and ± 2 times the standard error in the baseline respectively. The orange lines show models of fission fragments added to the baseline pattern; the dotted line has equal contributions from the baseline and the fission model, the dashed line has 2 parts fission fragments plus 1 part baseline pattern, and the solid line has 4 parts fission plus 1 part baseline. (**A**) Elements $34 < Z < 52$, normalized to Zr (filled circle). Elements are labeled at the bottom of the panel. (**B**) Residuals between the data and the baseline pattern in panel A. (**C & D**) Same as panels A and B, but for elements $56 < Z < 78$, normalized to Ba (filled circle). Numerical values are provided in Data S1.

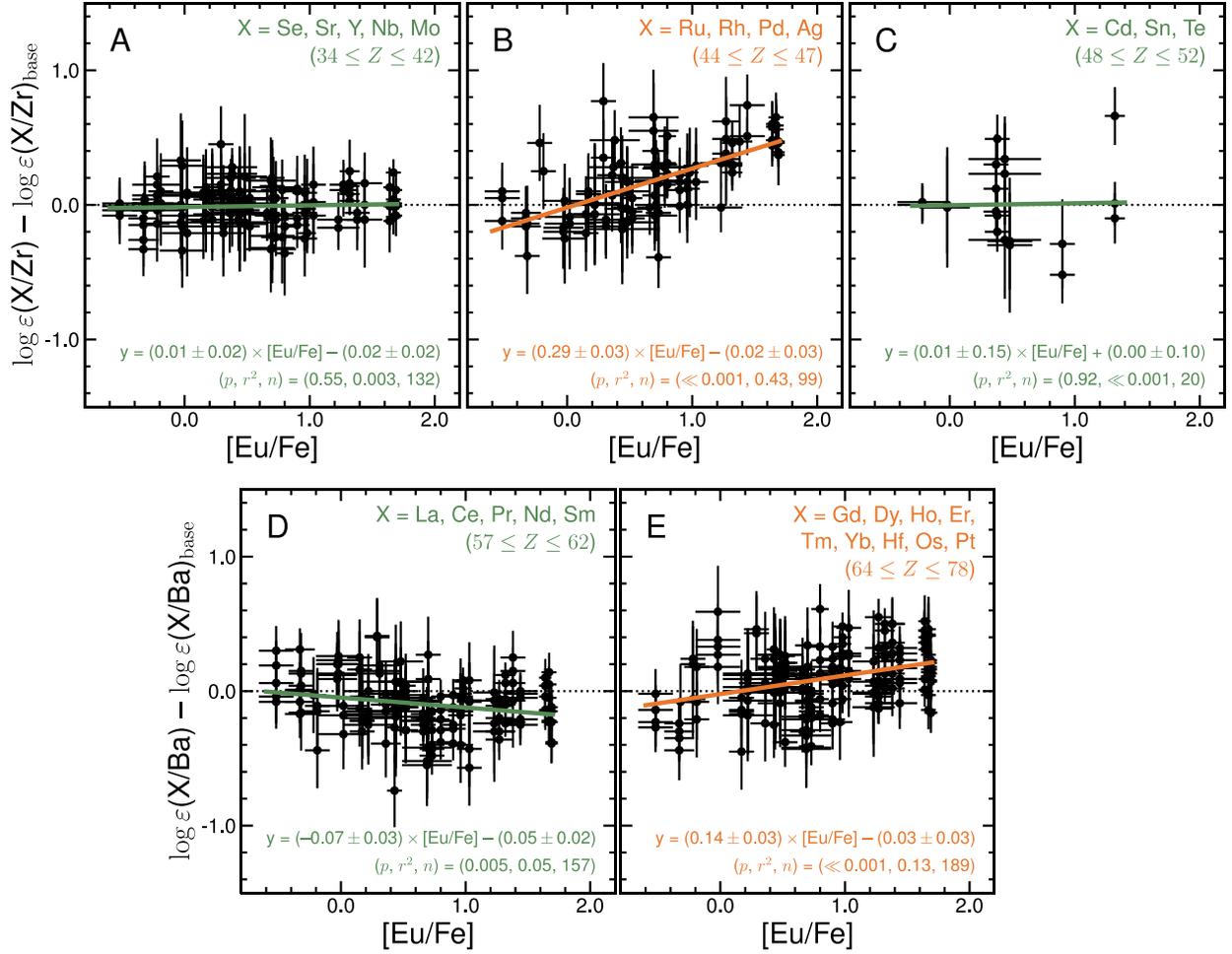

**Fig. 2: Abundance ratios of groups of elements that do or do not correlate with [Eu/Fe].** Each panel shows a different group of elements, ordered by increasing atomic number, $Z$: (**A**) $34 \leq Z \leq 42$, (**B**) $44 \leq Z \leq 47$, (**C**) $48 \leq Z \leq 52$, (**D**) $57 \leq Z \leq 62$ and (**E**) $64 \leq Z \leq 78$. In all panels, black data points are for one element abundance ratio in one star, with the error bars indicating $1\sigma$ uncertainties. Abundances have been normalized to Zr (panels A-C) or Ba (panels D-E), and the empirical baselines shown in Figure 1 have been subtracted. Solid lines show ordinary least-squares linear fits to the data in each panel, and dotted lines indicate differences of zero. Green indicates flat trends (lines consistent with zero); we interpret these as indicating that those elements are produced together with Zr or Ba. Orange indicates correlations that are significantly ($> 4.6\,\sigma$) different from zero; we interpret these as indicating those elements are produced together with Eu. Labels in each panel indicate the equation of each fitted line, the $p$-value of its Pearson correlation coefficient, the $r^2$ coefficient of determination, and the number of stars, $n$ (*13*); these values are also listed in Table S1. Separate plots for each element are shown in Figs. S3-S5.

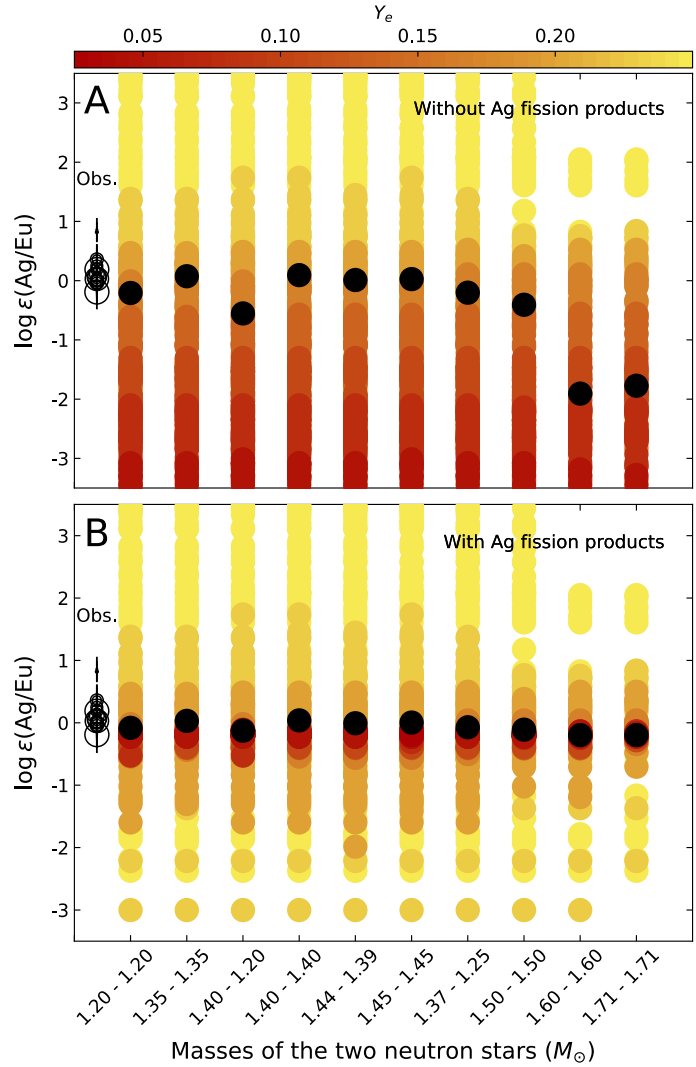

**Fig. 3: Ag/Eu abundance ratios predicted by nucleosynthesis models.** Predictions of log ε(Ag/Eu) from hydrodynamic simulations (*13*) of neutron star mergers are plotted as a function of the pair of progenitor masses. Overlapping colored dots show individual ejecta, while black dots are the mean mass-weighted abundance ratio (*13*). The color bar indicates the neutron richness $Y_e$ of individual ejecta; lower $Y_e$ ejecta are expected to experience more fission. Empty circles indicate the observed ratios in our stellar sample, with sizes proportional to the [Eu/Fe] ratio in each star (as in Fig. 1). Results are shown for models (**A**) without including fission and (**B**) with fission included. When fission is included, we find the predicted mass-weighted log ε(Ag/Eu) ratios are independent of the progenitor masses and are consistent with the observations.

**Acknowledgements:** We thank Ani Aprahamian, Eric Bell, Benoit Côté, Yutaka Hirai, Grant Mathews, Chris Sneden, and Michael Wiescher for comments on early versions of the manuscript.

**Funding:** We acknowledge support from the US National Science Foundation grants AST 1716251 (to AF); AST 1815403 (IUR); AST 1815767 (IUR); AST 2205847 (IUR); OISE-1927130 (IReNA: The International Research Network for Nuclear Astrophysics; TCB, AF, TTH); PHY 14-30152 (JINA-CEE: Joint Institute for Nuclear Astrophysics - Chemical Evolution of the Elements; IUR, EMH, MRM, RS, TCB, RE, AF); and PHY 2020275 (RS). RS acknowledges support from the US Department of Energy grants DE-FG02-95-ER40934 and DE-SC0018232. We acknowledge additional support from NASA grants 80NSSC21K0627 (Astrophysics Data Analysis Program; to IUR); HST-GO-15657 (IUR, TCB, AF, VMP); HST-GO-15951 (IUR, TCB, RE, AF, TTH, CMS); and HST-HF2-51481.001 (National Hubble Fellowship Program; EMH). NV acknowledges support from the Natural Sciences and Engineering Research Council of Canada. MRM acknowledges support from Los Alamos National Laboratory's Directed Research and Development program 20200384ER. TTH acknowledges support from grant VR 2021-05556 from the Swedish Research Council. VMP acknowledges support from NOIRLab, which is managed by the Association of Universities for Research in Astronomy under a cooperative agreement with the US National Science Foundation.

**Author contributions:** Conceptualization: IUR. Formal Analysis: IUR, TCB. Methodology: IUR, NV, EMH, MRM, RS. Investigation: IUR, NV. Visualization: IUR, NV, EMH, MRM, RS, JJC, TCB. Writing – original draft: IUR, NV. Writing – review & editing: IUR, NV, EMH, MRM, RS, JJC, TCB, RE, AF, TTH, VMP, CMS.

**Competing interests:** The authors declare that they have no competing interests.

**Data and materials availability:** Our compilation of stellar abundance data for the 42 stars in our sample is available in Data S1. The nucleosynthesis yields *(21)* we used as input for our fission calculations are archived at Zenodo *(27)*. Our derived correlations are quantified in Table S1 and the empirical r-process baseline abundances are listed in Tables S2 and S3.

**Supplementary Materials**
Materials and Methods
Supplementary Text
Figs. S1 to S6
Tables S1 to S3
Data S1
References (*28-130*)

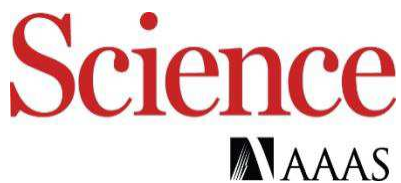

Supplementary Materials for

# Element abundance patterns in stars indicate fission of nuclei heavier than uranium

Ian U. Roederer*, Nicole Vassh, Erika M. Holmbeck, Matthew R. Mumpower,
Rebecca Surman, John J. Cowan, Timothy C. Beers, Rana Ezzeddine, Anna Frebel,
Terese T. Hansen, Vinicius M. Placco, Charli M. Sakari

*Corresponding author. Email: iuroederer@ncsu.edu

**This PDF file includes:**

Materials and Methods
Supplementary Text
Figs. S1 to S6
Tables S1 to S3
Caption for Data S1

**Other Supplementary Materials for this manuscript:**

Data S1 (.csv file)

**Materials and Methods**

Stellar sample

We assembled our stellar sample from previously published abundance measurements. Our sample is drawn from stars listed in the Joint Institute for Nuclear Astrophysics stellar abundance database (JINAbase) (*28*), supplemented with additional studies not included in JINAbase. We required that Zr, Ba, and Eu abundances are reported (and detected) for each star. We required that [Ba/Eu] < −0.3 for each star, which indicates that r-process material dominates the heavy-element abundance pattern (*29*). This cut excluded stars with contributions from other nucleosynthesis processes, such as the s-process or i-process. It does not affect our interpretation of a known deviation from r-process universality (variations in the [Ba/Eu] ratio), as discussed below. We also required that an abundance of at least one of the elements Se, Pd, or Te is reported, because we are interested in studying the abundances of elements in the mass range from A = 80 to 130. We included all the stars we could find that met those requirements.

Our sample is comprised of 42 stars spanning ≈ 2.6 dex in metallicity (−3.57 ≤ [Fe/H] ≤ −0.99) and ≈ 2.2 dex in r-process enhancement (−0.52 ≤ [Eu/Fe] ≤ +1.69). Data S1 lists the selected stars, their metallicity ([Fe/H]), r-process enhancement ([Eu/Fe]), elemental abundances and their uncertainties, and the references used. The uncertainties include both the statistical and systematic uncertainties, including the systematic uncertainties from the choice of model atmosphere parameters. Whenever the original reference did not explicitly state the abundance uncertainty, we calculated them based on the line-to-line abundance scatter and the data quality described in the original references (see below).

The elemental abundances of some stars were adopted from more than one literature source. In these cases, all abundances had been derived using the same set of model parameters. Often multiple references were by the same research group and adopted a consistent model. When abundances were adopted from multiple groups, we confirmed that the same model parameters were used. All abundances were derived using one-dimensional hydrostatic model atmospheres calculated under the assumption of local thermodynamic equilibrium (LTE). Radiative transfer calculations also assumed that LTE holds in the line-forming layers of the atmosphere.

We homogenized the stellar abundances from the literature as follows. Stellar abundances depend directly on the atomic transition probability of each line, expressed as the $\log(g_i f)$ value, where $g_i$ is the degeneracy of the lower level and $f$ is the oscillator strength. Different stellar abundance studies sometimes adopt different sets of $\log(g_i f)$ values for the same set of transitions, which artificially increases the statistical scatter in the set of combined abundances. Laboratory measurements of the quantities necessary to calculate $\log(g_i f)$ values of some of the lighter r-process elements have improved since the abundance measurements were published, which could improve the accuracy and precision of the stellar abundances. We therefore translated the abundances of Se through Te to a uniform $\log(g_i f)$ scale (*9*). The resulting shifts are always small, ≤ 0.14 dex, and typically ≤ 0.03 dex, so the impact on our results is minor. The updated log ε(*X*) abundances are provided in Data S1. The heavier r-process elements used consistent laboratory measurements (*30*), so no corrections are made for the abundances of these elements.

We also assessed the impact of using inhomogeneous stellar parameters from the literature on our results. We derived a homogeneous set of stellar parameters (effective temperature, $T_{\text{eff}}$, and surface gravity, log *g*) for all stars in our sample. We calculated $T_{\text{eff}}$ using broadband photometry (*GP* and *BP* filters) from Gaia Data Release 3 (*31*) and metallicity-sensitive color-$T_{\text{eff}}$ relations (*32*). These color-$T_{\text{eff}}$ relations were calibrated separately for dwarf and giant stars, so we separated the sample at a log *g* value of 3.7 (*32*). We adopted reddening estimates, *E(B-V)*, from dust maps (*33*). Many of our stars are in front of the dust responsible for the total reddening of the Milky Way. We therefore assumed *E(B-V)* = 0.0 for stars closer than 70 pc (*34, 35*), or we calculated *E(B-V)* using a previously published method (*36*). We de-reddened the *GP* and *BP* photometry using extinction coefficients (*37*). We calculated log *g* values [(*38*), their equation 1], assuming a typical metal-poor star mass of 0.8 ± 0.1 solar masses ($M_{\odot}$) (*39, 40*). We used de-reddened Gaia *RP* magnitudes and bolometric corrections (*41*) instead of Johnson *V* magnitudes in a version of this equation appropriate for this bandpass. We adopted distances based on Gaia parallax measurements, with corrected zero points (*42*). The parallax uncertainties are < 10% for 37 of the 42 stars (88%) in our sample, so we simply calculated distances from their inverse parallaxes. We omitted the remaining 5 stars from the test described in this paragraph. We estimated the statistical uncertainties in $T_{\text{eff}}$ and log *g* by drawing $10^4$ resamples of the values input to the stellar atmosphere models, assuming a normal distribution with a standard deviation equal to the quoted uncertainty (*38*). The values in Data S1 reflect the median and standard deviation of these resamples. Fig. S1 shows the Gaia-based $T_{\text{eff}}$ and log *g* values for these 37 stars are consistent with the literature values, but have a systematic offset ($T_{\text{eff}}$ (Gaia) − $T_{\text{eff}}$ (literature) = +140 ± 15 K, 1 std. dev. = 93 K; log *g* (Gaia) − log *g* (literature) = +0.20 ± 0.07 dex, 1 std. dev. = 0.40 dex). We retain the literature values for our subsequent analysis.

Literature studies for six stars in our sample (BPS CS 31078-018, BPS CS 31082-001, HD 20, HD 107752, HD 115444, and HD 122563) reported their sensitivity to $T_{\text{eff}}$ and log *g* for the abundance for each element. We used these relations to compute the abundance offsets for each element measured in each of these six stars that result from the $T_{\text{eff}}$ and log *g* differences between the Gaia scale and the literature scale (Fig. S2). We find that the elements with evidence of fission fragment deposition are not systematically impacted differently from the elements with no such signature. The differences in the mean abundances between the two groups are consistent with zero: Δ log ε(*X*/Zr) = +0.030 ± 0.013 dex, and Δ log ε(*X*/Ba) = −0.014 ± 0.012 dex. Therefore, the correlations in Figure 2 are not a consequence of adopting abundances derived using the literature stellar parameters. We conclude from this test that the use of an inhomogeneous set of $T_{\text{eff}}$ and log *g* values does not artificially induce the evidence for fission fragment deposition.

Most of the literature studies did not present sufficient information to rescale the stars to a fully homogeneous $T_{\text{eff}}$ and log *g* scale.

Abundance correlations and non-correlations

Figures S3 and S4 show the relationships with the [Eu/Fe] ratios for each of the lighter r-process elements (34 ≤ *Z* ≤ 52) and the heavier r-process elements (56 ≤ *Z* ≤ 78), respectively. Ordinary (unweighted) linear least-squares fits (*43*) are used.

Figure S3 shows no correlations between the log ε(*X*/Zr) ratios and [Eu/Fe] for most elements, *X*, where *X* = Se, Sr, Y, Nb, Mo, Cd, Sn, and Te. The slopes of these relations are not significantly different from zero by 3 times their uncertainties ($\sigma_m$). Using weighted linear least-squares fits, with weights proportional to the inverse-square uncertainties, does not change our conclusions, so we prefer the simpler unweighted fits. The *p*-value for the Pearson correlation coefficient was calculated using the Wald Test with a t-distribution of the test statistic (*43*); we cannot reject the null hypothesis of zero slope at even modest significance (we find $p \geq 0.08$, whereas $p < 0.05$ would indicate significance) for these elements. In contrast, the slopes are $> 3\sigma_m$ significant (and $p \leq 0.002$) for $X$ = Ru, Rh, Pd, and Ag.

Fig. S4 shows the slopes between the log ε(*X*/Ba) ratios and [Eu/Fe], which are not significant (at the $3\sigma_m$ level) for $X$ = La, Pr, Nd, Sm, Tb, Ho, Er, Tm, Yb, Hf, Os, and Pt. Significant positive slopes ($\geq 3\sigma_m$) are found for $X$ = Gd and Dy (and $p \leq 0.008$), and a significant negative slope ($> 3\sigma_m$) is found for $X$ = Ce ($p = 0.006$). Heavier r-process elements with more limited numbers of measurements (Tb, Ho, Er, Tm, Yb, Hf, Os, Pt) also have positive correlations, but their slopes are not significant at the $3\sigma_m$ level. The use of weighted linear least-squares fits does not change these conclusions; though it does increase the significance for $X$ = Er and Yb (weighted slopes = 0.15 ± 0.05 and 0.24 ± 0.06, respectively). The significant ($>3\sigma_m$) negative slope (−0.13 ± 0.04) for Ce ($Z = 58$) is in the opposite sense as the correlations we associate with signatures of fission fragments. We have not identified any potential cause of this negative correlation.

We used these results for individual elements to collate them into the five groups, shown in Fig. 2: $34 \leq Z \leq 42$ (each have no significant correlations with [Eu/Fe]), $44 \leq Z \leq 47$ (each have significant positive correlations with [Eu/Fe]), $48 \leq Z \leq 52$ (no significant correlations with [Eu/Fe]), $56 \leq Z \leq 62$ (no significant positive correlations with [Eu/Fe]), and $64 \leq Z \leq 78$ (positive correlations, some significant and some not, with [Eu/Fe]).

We next assessed the statistical significance of the correlations for each of these five element groups. The slopes of the groups shown in Fig. 2A and 2C are $< 1\sigma_m$ significant, and the slope of the group shown in Fig. 2D is negative and consistent with zero within $2.3\sigma_m$. The *p*-value for the Pearson correlation coefficient for each sample cannot reject the null hypothesis of zero slope at even modest significance ($p \geq 0.05$; Fig. 2A and 2C). The *p*-value for the third set (Ba through Sm; Fig. 2D) is 0.005 for 157 measurements. The $r^2$ (coefficient of determination) values (*43*) for the elements shown in Fig. 2A, 2C, and 2D indicate that 0.3%, $\ll 0.1\%$, and 5%, respectively, of the variations in the abundances can be associated with the correlation with [Eu/Fe], rather than random uncertainties in the abundances. We conclude that the elements shown in Fig. 2A, 2C, and 2D do not exhibit significant ($> 3\sigma_m$) positive correlations with [Eu/Fe].

The groups shown in Fig. 2B and 2E exhibit different behavior. Their slopes differ from zero by more than $4.6\sigma_m$. Their *p*-values are highly significant (Ru through Ag: $p \ll 0.001$ for $n = 99$ individual measurements; Gd through Pt: $p \ll 0.001$ for $n = 189$ individual measurements). Their $r^2$ values indicate that 43% and 13%, respectively, of the variations in the abundances can be associated with the correlation with [Eu/Fe]. These metrics collectively indicate that the relationships shown in Fig. 2B and 2E are statistically significant.

There are no significant correlations between the abundance ratios log ε(Ru/Zr), log ε(Rh/Zr), log ε(Pd/Zr), log ε(Ag/Zr) and either $T_\text{eff}$ or log $g$ among the stars in our sample. These relationships are shown in Fig. S5. Specifically, the *p*-values for these correlations are all high, $\geq 0.09$, and the $r^2$ values are all low, indicating that less than 15% (and frequently less than 2%) of the variations in the abundances can be associated with the correlations with $T_\text{eff}$ or log $g$. We conclude that the newly identified abundance behavior is not a consequence of systematic trends related to the stellar parameters.

We assumed that the abundance uncertainties are normally distributed, because most of the spectral lines of those elements are "weak" in these stars, in the sense that they exhibit an approximately linear relationship between the line strength and the abundance in the line-forming layers of the atmosphere. Most of these lines also respond linearly to small changes in the stellar parameters, in the sense that a typical uncertainty in $T_{\text{eff}}$, log $g$, or microturbulent velocity parameter induces a linearly scaled uncertainty in the derived abundance (*44*).

We also assumed that the abundances of different elements are independent of each other, because the spectral lines of each element are generally unblended (*45, 46*). The behavior of an unblended line does not depend on the abundances of other r-process elements, because the abundances of the r-process elements are negligible when calculating the model atmosphere or continuous opacity. The use of abundance ratios minimizes the impact of correlated abundances that could result from stellar parameter uncertainties, as discussed above.

Nucleosynthesis calculations: hydrodynamics, nuclear data, and fission model

We adopted the yields from nucleosynthesis calculations performed previously (*21*) based on the output of previous hydrodynamic simulations of binary neutron star merger dynamical ejecta (*23*). Those were performed for several variations of the neutron star equation of state, different treatments of neutrino transport, and different sets of progenitor masses. We adopted the results from the case of the LS220 equation of state with a neutrino leakage scheme, because those were used for the widest set of progenitor masses (*23*). Our adoption of these models does not imply that neutron star mergers are the only site of r-process nucleosynthesis. Rather, we use this scenario to explore a broad set of conditions for fission of transuranic nuclei in an r-process environment.

Figure 3 shows the stabilizing effect that fission deposition has on the predicted abundance ratios. We considered two distinct fission yield sets for neutron-rich nuclei. In the first, it was assumed that a species always splits in half during fission, concentrating products near $A \sim 130$ and producing no lighter elements, such as Ag. In the second, a theoretical model for the fission yields was assumed (*47*), which uses predictions from the finite range liquid drop model (FRLDM) (*48*) for very heavy unstable species. This model produces substantial fission products with $A < 130$. Although some previous studies used theoretical fission yield models that did not have large contributions to r-process abundances below $A = 130$ (*17*), other models do predict substantial contributions to the light heavy elements via asymmetric yield distributions with a lighter and heavier peak (*25, 49, 50*).

The results shown in Fig. 3 reflect the mass-weighted abundance ratio. This ratio is defined as

$$\Sigma_i \frac{m_i}{M_{\text{tot}}} Y(\text{Ag})_i / \Sigma_i \frac{m_i}{M_{\text{tot}}} Y(\text{Eu})_i \,, \tag{S1}$$

where $M_{\text{tot}}$ is the total mass of all ejecta, and $Y_i$ are the abundances predicted for the individual ejecta components of mass $m_i$ for Ag and Eu.

To explore variations in the fission yields, we kept all other input nuclear data the same and adopted other existing datasets (*21*). These included mass predictions from the Finite Range Drop Model (FRDM2012) (*51*), with neutron capture and neutron-induced fission rates determined from the Los Alamos National Laboratory (LANL) Coupled channels and Hauser-Feshbach (CoH) model (*52*), and β-decay and β-delayed fission rates determined from the Beta-decay by CoH (BeoH) (*53*) model. We also adopted the Finite Range Liquid Drop Model (FRLDM) fission barriers (*54*). Although large uncertainties remain in the nuclear data for the neutron-rich nuclei populated during these nucleosynthesis calculations, which affect the exact abundance ratios predicted (*49*), our identification of fission deposition and coproduction does not depend on our choice of adopted nuclear data.

Modeling the abundance patterns with the inclusion of fission fragments

We calculated how the abundance pattern changes when including fission fragments. Our empirical template for the baseline abundance pattern was calculated using the mean abundance ratios found in the 13 stars with [Eu/Fe] ≤ +0.3. We assumed that these stars contain minimal contributions from whatever process is responsible for producing the abundance excesses observed in Figs. 1 and 2. The star-to-star dispersion of the abundance ratios in these 13 stars (median σ of 0.19 dex) is similar to the typical observational uncertainties (median uncertainty of 0.22 dex), so could be entirely due to observational uncertainties.

Tables S3 and S4 list this baseline pattern, expressed as log ε($X$/Zr)$_{base}$ and log ε($X$/Ba)$_{base}$, which is shown in Fig. 1. We assumed that this baseline r-process pattern is present in all stars in the sample. We added enhancements of transuranic fission fragments, adopted from the models described in the previous section. We scaled the level of fission fragment enhancement, $F_{enh}$, to the ratio log ε(Pd/Zr) for the lighter r-process elements and the ratio log ε(Gd/Ba) for the heavier r-process elements. For example, $F_{enh}$ = 2 indicates 1 part baseline and 2 parts fission fragments. Mathematically, the abundance ratio for element $X$ relative to Zr can be expressed as

$$\log \varepsilon (X/\text{Zr})_{\text{total}} = \log(F_{\text{enh}} \times 10^{\log \varepsilon(X/\text{Zr})_{\text{fiss}}} + 10^{\log \varepsilon(X/\text{Zr})_{\text{base}}}), \qquad (S2)$$

where

$$\log \varepsilon (X/\text{Zr})_{\text{fiss}} = \log \varepsilon (X)_{\text{fiss}} - \log \varepsilon (\text{Pd})_{\text{fiss}} + \log \varepsilon (\text{Pd/Zr})_{\text{base}}. \qquad (S3)$$

The values of log ε($X$)$_{fiss}$ and log ε(Pd)$_{fiss}$ were adopted from the fission model with equal-mass 1.40 M$_\odot$ merging neutron stars, although the exact model selected has little influence on the result, as shown in Fig. 3. All logarithms are base 10. We used an analogous set of equations to describe the enhancements relative to the baseline abundance pattern for the heavier r-process elements, replacing Zr with Ba and Pd with Gd. The enhancements described by these equations are shown in Fig. 1.

We estimated the relative contributions from fission fragment deposition under the assumption that stars with [Eu/Fe] ≤ +0.3 contain minimal contributions from fission. Using the weighted least-squares linear fit shown in Fig. 2, we find the Ru through Ag abundances increase by a factor of ≈ 2 (0.30 dex) in stars with [Eu/Fe] ≈ +1.1. Thus, roughly half of the Ru through Ag in stars with [Eu/Fe] ≈ +1.1 originated as fission fragments. These elements are enhanced by an average factor of ≈ 3 (0.48 dex) in the stars with the highest r-process enhancement, indicating that fission could be responsible for up to ≈ 75%, of the elements from Ru to Ag in these stars.

We also estimated the approximate fraction of metal-poor stars where fission fragment deposition contributes substantially to the r-process element inventory. The R-Process Alliance (RPA) sample (*55-58*) includes 595 metal-poor stars, approximately 97% of which have $-3.0 \leq$ [Fe/H] $\leq -1.5$. This sample is biased in favor of r-process-enhanced stars. Of the RPA targets, 304 stars (51%) exhibit [Eu/Fe] ratios $\geq +0.3$ (*58*). In other words, fission fragment deposition could have impacted up to half of the stars in this metallicity range, so is a potentially common phenomenon.

**Supplementary Text**

Previous investigations of deviations from r-process abundance universality

R-process universality is usually assumed to apply to heavy r-process elements, including those with $56 \leq Z \leq 72$ (*45*) and $76 \leq Z \leq 78$ (*29*). Two deviations from r-process universality have been identified in previous studies. The larger effect of the two, which spans more than 1.5 dex from one metal-poor star to another, is characterized by differing abundance levels between the lighter ($34 \leq Z \leq 52$) and heavier ($Z \geq 56$) r-process elements (*59-61*). The abundances of the two element groups are not independent of each another. The abundance pattern within each of the two groups is generally consistent, once the overall abundance level has been normalized (*45, 62, 63*), except for the fission excesses we have identified. The second effect is smaller, with variations up to 0.7 dex; it is characterized by differing abundance levels between the actinides (Th and U; $Z = 90$ and 92) and the heaviest stable r-process elements (*64, 65*). This effect is known as the actinide boost (when the actinides are enhanced) and is discussed below. These deviations from universality could reflect a diversity of conditions within a given r-process site (*66, 67*), multiple sites within individual events (*68-71*), multiple events (*15*), or some combination of these scenarios. Neither of these deviations is related to the deviations we have identified.

Other evidence for cosmic production of transuranic nuclei

Previous studies have found evidence for cosmic production of the transuranic nuclei up to at least $A = 247$. The decay products of $^{244}$Pu and $^{247}$Cm have been found in meteorites (*72, 73*) and $^{244}$Pu has been directly detected in seafloor sediments (*74, 75*). Transuranic nuclei have not been observed in stars, perhaps because their half-lives are short ($\leq$ 80 Myr) compared to the ages of stars in which they might otherwise be observable ($\gg$ 1 Gyr).

Alternative explanations

The correlations between the ratios log ε(Ru/Zr), log ε(Rh/Zr), log ε(Pd/Zr), and log ε(Ag/Zr) and [Eu/Fe] indicates that the source of the elevated Ru, Rh, Pd, and Ag abundances is related to the source of the heavy elements, including Eu. In principle, that source could be a different nucleosynthesis process operating in a core-collapse supernova, such as the weak r-process (*76*), weak s-process (*77*), or the intermediate neutron-capture process (i-process) (*78*). We exclude any process that operates only in low- or intermediate-mass stars that pass through the asymptotic giant branch (AGB) phase of evolution, such as the main s-process (*79*), because these sites produce little Eu compared with other heavy elements. Therefore, we compare the observed abundance behavior with predictions for the weak r-process, the weak s-process, and the i-process.

The weak r-process could be achieved in environments such as core-collapse supernovae and accretion-disk ejecta, where matter is only slightly neutron rich. The weak r-process is not predicted to produce substantial amounts of elements as heavy as Eu (*76, 80, 81*). Therefore, the weak r-process is not a viable explanation for the observed abundance behavior.

The weak s-process occurs in massive, rapidly rotating stars. Models of the weak s-process in low-metallicity environments (*77, 82, 83*) do not predict enhancement among any of the ratios log ε(Ru/Eu), log ε(Rh/Eu), log ε(Pd/Eu), or log ε(Ag/Eu) without a similar increase in the log ε(Ba/Eu) and log ε(Pb/Eu) ratios. These models do predict enhancement of the log ε(Ba/Eu) and log ε(Pb/Eu) ratios by several orders of magnitude, which exceed the observed ratios: log ε(Ba/Eu) = 0.99 ($\sigma$ = 0.07 dex) in the nine stars with [Eu/Fe] > +1.0 in our sample, and log ε(Pb/Eu) ≈ 0.8 ($\sigma$ ≈ 0.2 dex) in a previous study of other r-process-enhanced stars (*84*). Therefore we exclude the weak s-process as a viable explanation for the observed abundance behavior.

The i-process has been associated with several sites, including massive stars, super-AGB stars, post-AGB stars, He-core and He-shell flashes in low-mass stars, and rapidly accreting white dwarfs. Models of the i-process produce both Pd and Eu, as well as other heavy elements (*85-88*). Models that produce log ε(Pd/Eu) ratios similar to those observed in the nine stars with [Eu/Fe] > +1.0, log ε(Pd/Eu) = 0.71 ($\sigma$ = 0.12 dex), also predict log ε(Ba/Eu) ratios $\gtrsim$ 2. That is much higher than observed in these nine stars, log ε(Ba/Eu) = 0.99 ($\sigma$ = 0.07 dex). Furthermore, the i-process would need to dominate the production of Eu in the stars with the highest levels of [Eu/Fe], which is unlikely given that the abundance ratios among Eu, the lanthanide elements, and heavier elements up to $Z$ = 79 are consistent with the Solar System r-process (*9, 89*). Therefore, we also exclude the i-process as a viable explanation for the observed abundance behavior.

We conclude that the only known nucleosynthesis process that can explain the observed abundance behaviors is deposition of transuranic fission fragments during the r-process.

Using [Ba/Eu] to assess contributions from the r- and s-process

A high percentage (> 94%) of the Eu in the Solar System originated via r-process nucleosynthesis (*90-92*). In contrast, a high percentage of the Ba in the Solar System (> 85%) originated via s-process nucleosynthesis (*90-92*). We therefore use the [Ba/Eu] ratio [or, analogously log ε(Ba/Eu)] as a metric to assess the relative contributions of the r- and s-process to stars. The exact values expected for pure r- and s-process [Ba/Eu] ratios depend on the adopted model and method, but are typically separated by > 2 dex (*91, 93, 94*). For example, one study (*92*) estimated that the [Ba/Eu] ratios for pure r-process and s-process material are ≈ −0.9 and ≈ +1.3, respectively.

Our findings imply that the [Ba/Eu] ratio exhibits a small amount of cosmic dispersion due to the r-process itself. The eight stars in our sample with [Eu/Fe] < 0.0 have [Ba/Eu] = −0.53 ± 0.05 (log ε(Ba/Eu) = 1.13 ± 0.05), while the nine stars with [Eu/Fe] > +1.0 have [Ba/Eu] = −0.67 ± 0.02 (log ε(Ba/Eu) = 0.99 ± 0.02). These differences are small and about 2.5σ significance.

We therefore regard the [Ba/Eu] ratio as a general diagnostic of the r- and s-process contributions. The differences between the low- and high-[Eu/Fe] samples are much smaller than the > 2 dex differences between pure r- and s-process abundance ratios. Caution is warranted, however, when interpreting small (< 0.2 dex) enhancements in the [Ba/Eu] ratio, which might not necessarily imply the presence of small amounts of s-process contamination in an otherwise dominant r-process pattern. Similar caution should be applied to other element ratios, such as [La/Eu] (*95*) or [Ce/Eu], where small but measurable enhancements could also signal deficiencies in Eu that result from less fission fragment deposition.

A metric to assess the strength of the r-process

A previous modelling study (*21*) compared the predictions from their nucleosynthesis calculations with a limited set of observational data in the form of 13 r-process-enhanced stars drawn from JINABase (*28*). That study examined the relationships between the [Ru/Eu], [Pd/Eu], and [Ag/Eu] ratios and the log ε(Eu) ratio, which they adopted as a measure of the level of r-process enrichment. They found no correlations between the [Ru/Eu], [Pd/Eu], or [Ag/Eu] ratios and log ε(Eu), which was consistent with the behavior expected if the Ru, Pd, and Ag originate in part as fission fragments. We performed a similar analysis for the stars in our sample, finding no significant correlations between log ε(Eu) and each of log ε(Ru/Zr) (slope = 0.13 ± 0.06, $p$ = 0.03, 29 stars) and log ε(Ag/Zr) (slope = 0.14 ± 0.06, $p$ = 0.02, 21 stars). These correlations are weaker than the correlations with [Eu/Fe].

We find that [Eu/Fe], not log ε(Eu) (or [Eu/H]), is the metric that most closely relates to the strength of the r-process. It is unclear why this is so. The log ε(Eu) ratios reflect the amount of Eu and r-process elements present in the gas from which these stars formed. In contrast, the [Eu/Fe] ratios reflect the amount of Eu and r-process elements present relative to the Fe abundance of the gas, so they depend on the metallicity of the environment where the r-process occurred (*11, 96, 97*).

Comparison with the actinide boost phenomenon

Figure S6A shows the actinide boost phenomenon (discussed above) is not correlated with the production of transuranic fission fragments in our sample. We adopt log ε(Th/Eu) as a measure of the range of actinide production relative to the lanthanide elements, and log ε(Pd/Zr) as a measure of the fission fragment yields relative to the lighter r-process elements. For the 15 stars in the sample with reported Pd and Th abundances, the $p$-value for the Pearson correlation coefficient between the log ε(Th/Eu) and log ε(Pd/Zr) ratios is 0.90, indicating no significant correlation. Furthermore, the variation in the observed actinide abundances relative to the lanthanide abundances is much smaller than the variation observed among the fission fragments. The ranges of the log ε(Th/Eu) and log ε(Pd/Zr) ratios are 0.45 dex and 1.13 dex, respectively, which correspond to factors of 2.8 and 13.5. The abundances of fission fragments vary by ≈ 5 times more than the actinide variations.

Figure S6B shows the relationship between the fission fragments and the actinides, which is also not correlated with the strength of the r-process. We adopt log ε(Pd/Th) as a measure of the amount of fission fragments relative to the actinides and [Eu/Fe] as a measure of the strength of the r-process. The log ε(Pd/Th) ratio is not correlated with the [Eu/Fe] ratio. The $p$-value for the Pearson correlation coefficient between the log ε(Pd/Th) and [Eu/Fe] ratios is 0.37 for the 15 stars in our sample, indicating no significant correlation.

We conclude that there is no correlation between the fission fragment abundances and the actinide abundances in these stars, so the actinide boost phenomenon is not a viable explanation for our results.

Limitations of our analysis

We omitted numerous heavy elements (Br, Kr, Rb, In, Sb, I, Xe, Cs, Lu, Ta, W, Re, Ir, Au, Hg, Tl, Pb, Bi, and U) from our analysis because there are insufficient observations of them available. Tc, Pm, Po, At, Rn, Fr, Ra, Ac, and Pa were also omitted because they have no stable or long-lived isotopes. Th is discussed separately above.

The baseline abundance pattern we derive (Fig. 1, Tables S3 and S4) is calculated empirically, not self-consistently with the fission component. Our analysis preserves the observed abundance ratios among Se, Sr, Y, Zr, Nb, Mo, Rh, and Pd (Fig. 1). This underpredicts the abundances of Ru and overpredicts the abundances of Ag, Cd, Sn, and Te, relative to the observations. This could indicate that only a narrow range of nuclei undergo fission during the r-process, which would narrow the range of daughter products produced.

Our conclusion that some elements partly originate from transuranic fission fragment deposition during the r-process might not apply to stellar populations that are contaminated by material from other processes, such as the s-process. A previous study (*98*) examined the abundances of Sr, Y, Zr, Pd, Ag, Ba, and Eu in 71 metal-poor stars and traced the contributions from different nucleosynthesis processes to each of these elements. They reached the opposite conclusion from our study, finding that Pd and Ag correlated more strongly with Zr than with Eu. The cause of this discrepancy is unclear, but could be related to contamination of material produced by other processes in the gas from which these stars formed. The two samples have overlapping ranges of stellar parameters, metallicities, and [Eu/Fe] ratios, but our sample includes stars with a more dominant r-process signature: the median [Ba/Eu] ratio for our sample is −0.55, while for the previous study (*98*) it was −0.21. The sample used in the previous study (*98*) was selected to include stars that would reflect a combination of nucleosynthesis processes, such as charged-particle nucleosynthesis, the weak r-process, the s-process, and the r-process. In contrast, our sample was selected to only include stars where the r-process is the dominant source of the heavy elements.

Observations of r-process elements in stars are limited to elemental abundances. Individual isotopic abundances are generally inaccessible, with only a few exceptions (*99, 100*). Our finding that Eu and Ba are slightly decoupled when transuranic fission fragment deposition occurs suggests that measurements of the Ba isotopes in very metal-poor stars (*101-104*) may need to be reinterpreted in terms of the nucleosynthetic origins of Ba in light of potential variations in the r-process [Ba/Eu] ratio, as discussed previously.

Not all elements can be detected in all stars. For example, there are very few observations of Cd, Sn, Te, Hf, Os, and Pt, which limits our ability to draw conclusions about their origins. This limitation arises from the conditions in stellar atmospheres and the energy spacing of those elements' electronic configurations. These factors restrict detection of these elements to ultraviolet wavelengths, which are observable only from space (*9*).

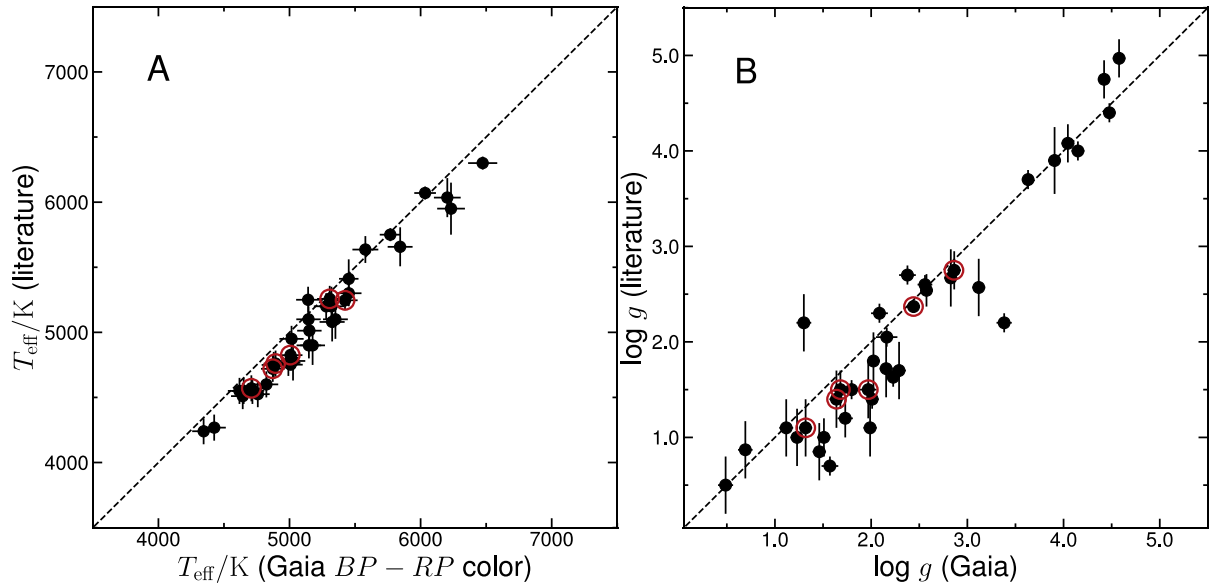

**Figure S1: Comparison of $T_{\rm eff}$ and log *g* values from the literature and using our analysis of Gaia data.** (**A**) $T_{\rm eff}$ values, and (**B**) log *g* values. The dashed line marks a 1:1 relation. Error bars indicate 1σ uncertainties (see text). The red circles mark six stars with abundance sensitivities listed in the literature, discussed in the text.

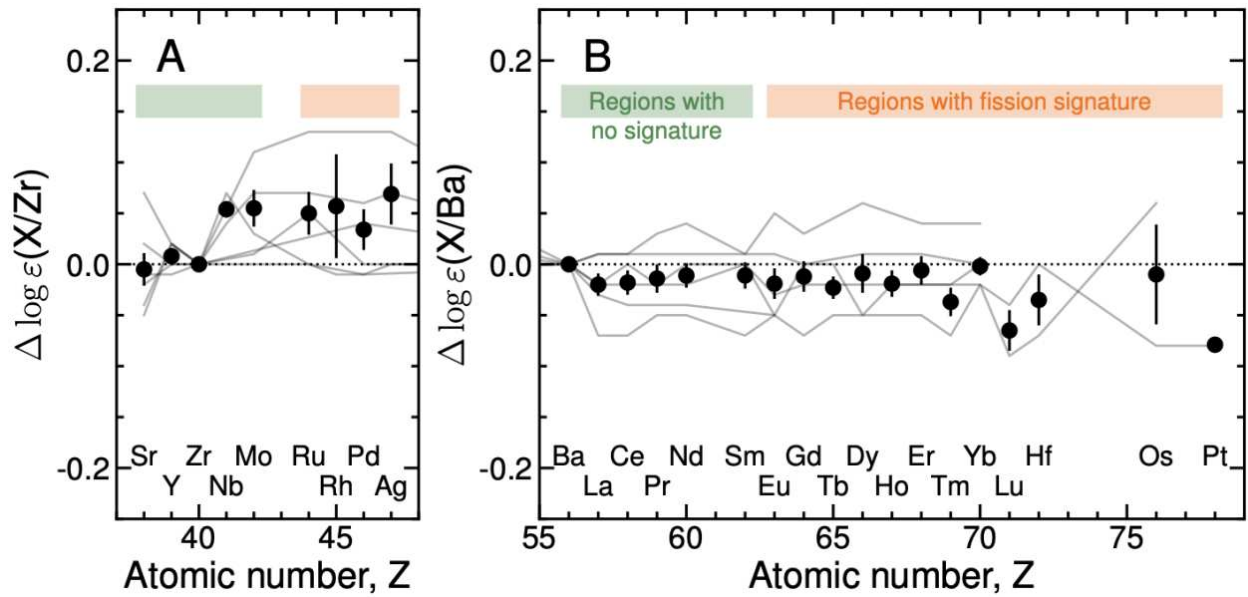

**Figure S2: Differences between abundance ratios from the literature and our analysis of Gaia data.** (**A**) the abundances of lighter elements, normalized to Zr. (**B**) the abundances of heavier elements, normalized to Ba. The gray lines show the measurements for individual stars, and the large black points show the mean (with error bars showing the standard error) of these offsets. The orange boxes mark elements in which we have identified the signature of fission fragment deposition, and the green boxes mark elements where we have not. The abundance differences are not systematically different between these two groups (see text).

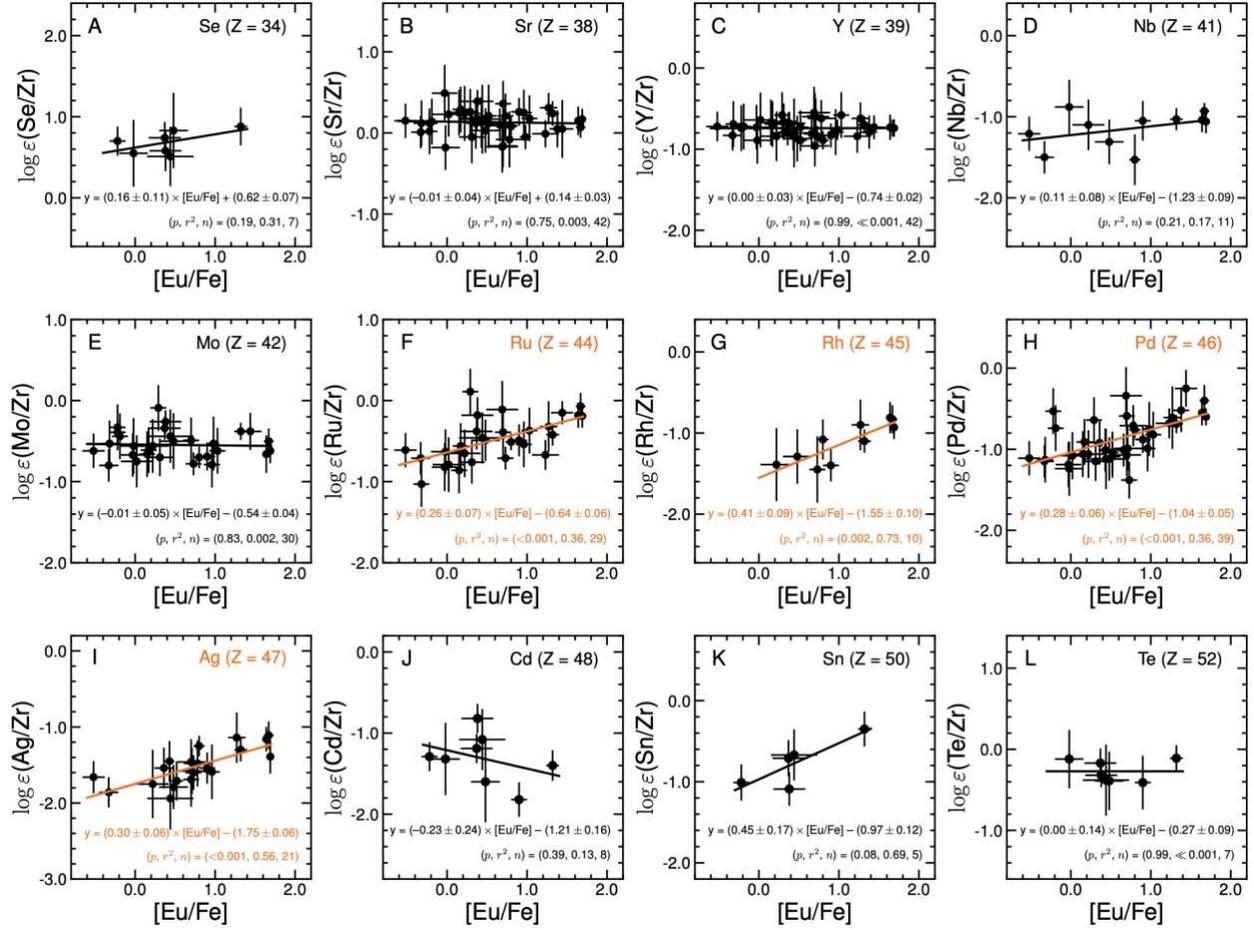

**Fig. S3: Abundance ratios among the lighter elements.** Each data point represents one star, with error bars representing 1σ uncertainties. All 42 stars are included, but not every star has a measurement for each element. Each panel illustrates the abundance ratios for a different element $X$ (where $X$ = Se, Sr, ..., Te) arranged by increasing atomic number, $Z$. Lines show ordinary least-squares linear fits, with the equations for these lines listed in each panel. Numerical values are provided in Table S1. Non-significant slopes (< 3σ; black lines) indicate that an element is coproduced with Zr, and significant positive correlations (≥ 3σ; orange lines) indicate that an element is coproduced with Eu. The three values labelled in the lower right corner of each panel are the $p$-value for the Pearson correlation coefficient, the $r^2$ coefficient of determination, and the number of stars, $n$.

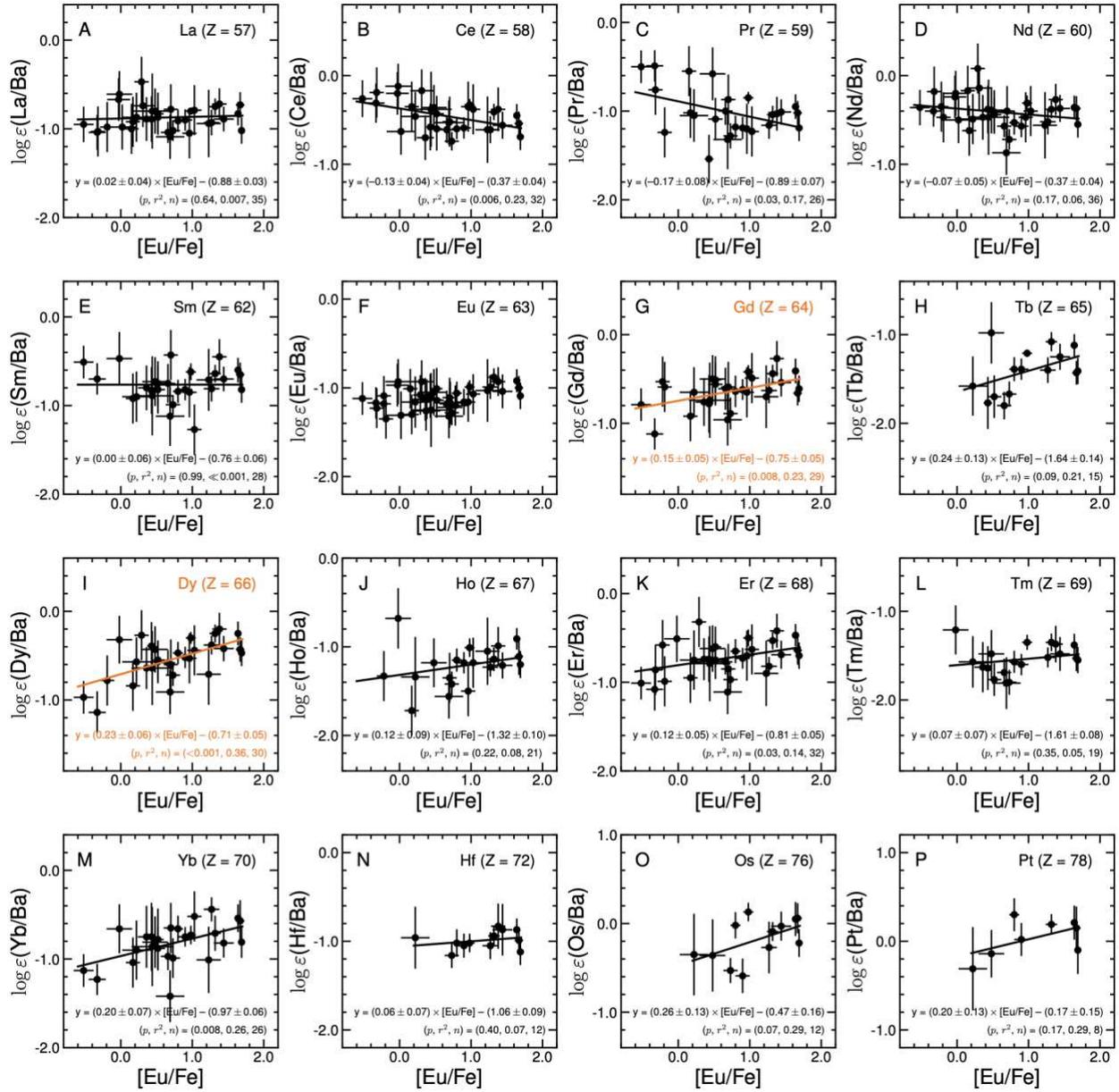

**Figure S4: Abundance ratios among the heavier elements.** Same as Fig. S3 but for (**A**-**M**) the lanthanides La, Ce, Pr, Nd, Sm, Eu, Gd, Tb, Dy, Ho, Er, Tm, Yb; and (**N**-**P**) the elements Hf, Os, Pt near and at the third r-process peak. No values are labelled in panel F, because the two variables are not independent.

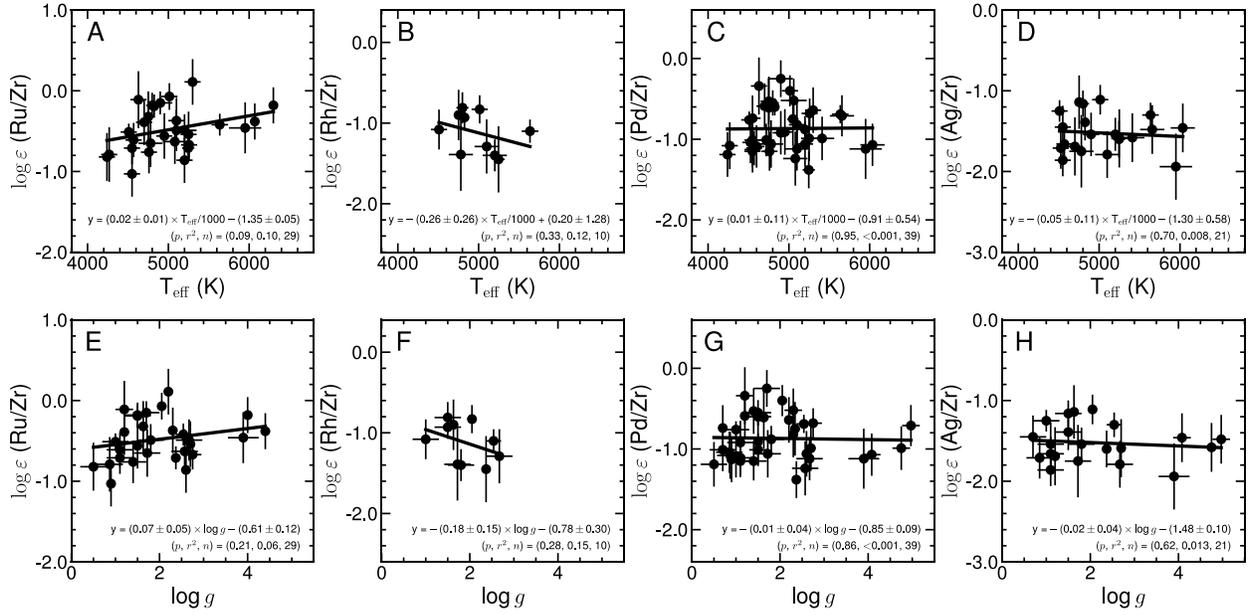

**Figure S5: Abundance ratios among Zr, Ru, Rh, Pd, and Ag as a function of the stellar parameters $T_{eff}$ and log $g$.** Plotting symbols and labels are the same as in Fig. S3. No significant correlations are found.

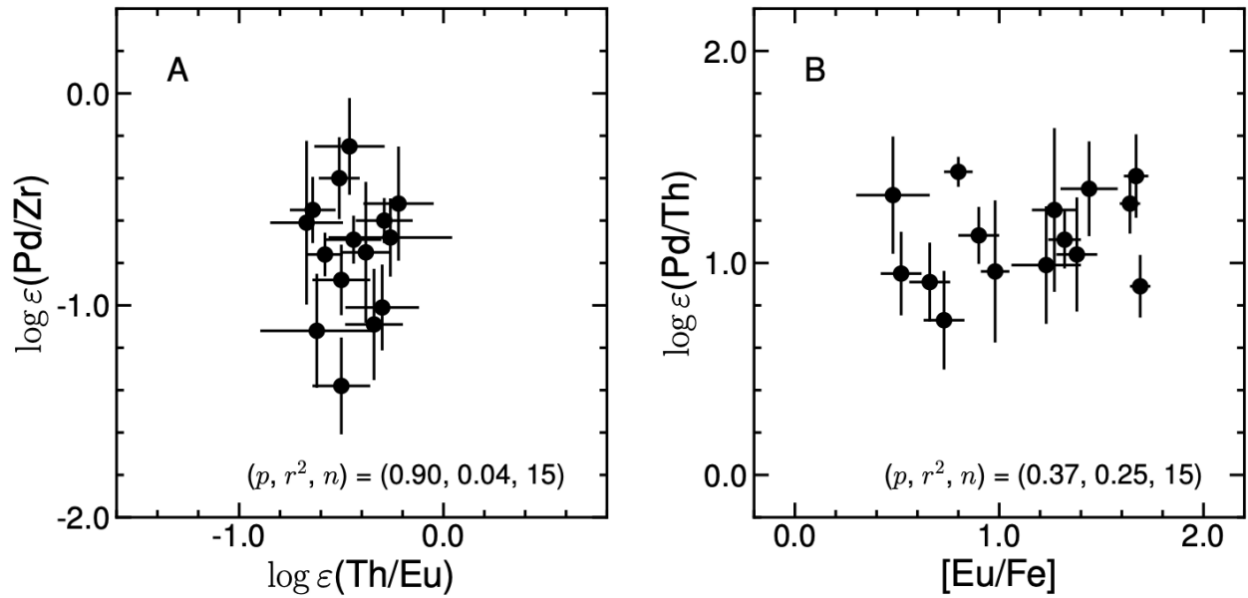

**Figure S6: Relationships between fission fragment abundances and actinides.** (**A**) the ratio log ε(Pd/Zr), representing the relationship between the fission fragments and lighter r-process elements, as a function of the ratio log ε(Th/Eu) representing the relationship between the actinides and the lanthanides. (**B**), the ratio log ε(Pd/Th) representing the relationship between the fission fragments and the actinides, as a function of the [Eu/Fe] ratios representing the strength of the r-process. All axes span 2.4 dex. Data points mark individual stars, and error bars are 1σ uncertainties. The three values labelled in each panel are the $p$-value for the Pearson correlation coefficient, the $r^2$ coefficient of determination, and the number of stars, $n$. Neither correlation is significant.

**Table S1. Best-fitting correlations between different abundance ratios and [Eu/Fe].** Column 1 specifies the elements used to calculate the ratio log ε(A/B) – log ε(A/B)$_{base}$. Column 2 lists the slope of the relation between the element ratio in column 1 and [Eu/Fe] and column 3 lists the intercept. Columns 4 and 5 list the $p$ value and $r^2$ value. Column 6 lists the number of stars, $n$, used in fitting each correlation.

| A/B | Slope | Intercept | $p$ | $r^2$ | $n$ |
|---|---|---|---|---|---|
| Se,Sr,Y,Nb,Mo/Zr | 0.01 ± 0.02 | −0.02 ± 0.02 | 0.55 | 0.003 | 132 |
| Ru,Rh,Pd,Ag/Zr | 0.29 ± 0.03 | −0.02 ± 0.03 | < 0.001 | 0.43 | 99 |
| Cd,Sn,Te/Zr | 0.01 ± 0.15 | 0.00 ± 0.10 | 0.92 | < 0.001 | 20 |
| La,Ce,Pr,Nd,Sm/Ba | −0.07 ± 0.03 | −0.05 ± 0.02 | 0.005 | 0.05 | 157 |
| Gd,Dy,Ho,Er,Tm,Yb,Hf,Os,Pt/Ba | 0.14 ± 0.03 | −0.03 ± 0.03 | < 0.001 | 0.13 | 189 |
| Se/Zr | 0.16 ± 0.11 | 0.62 ± 0.07 | 0.19 | 0.31 | 7 |
| Sr/Zr | −0.01 ± 0.04 | 0.14 ± 0.03 | 0.75 | 0.003 | 42 |
| Y/Zr | 0.00 ± 0.03 | −0.74 ± 0.02 | 0.99 | < 0.001 | 42 |
| Nb/Zr | 0.11 ± 0.08 | −1.23 ± 0.09 | 0.21 | 0.17 | 11 |
| Mo/Zr | −0.01 ± 0.05 | −0.54 ± 0.04 | 0.83 | 0.002 | 30 |
| Ru/Zr | 0.26 ± 0.07 | −0.64 ± 0.06 | < 0.001 | 0.36 | 29 |
| Rh/Zr | 0.41 ± 0.09 | −1.55 ± 0.10 | 0.002 | 0.73 | 10 |
| Pd/Zr | 0.28 ± 0.06 | −1.04 ± 0.05 | < 0.001 | 0.36 | 39 |
| Ag/Zr | 0.30 ± 0.06 | −1.75 ± 0.06 | < 0.001 | 0.56 | 21 |
| Cd/Zr | −0.23 ± 0.24 | −1.21 ± 0.16 | 0.39 | 0.13 | 8 |
| Sn/Zr | 0.45 ± 0.17 | −0.97 ± 0.12 | 0.08 | 0.69 | 5 |
| Te/Zr | 0.00 ± 0.14 | −0.27 ± 0.09 | 0.99 | < 0.001 | 7 |
| La/Ba | 0.02 ± 0.04 | −0.88 ± 0.03 | 0.64 | 0.007 | 35 |
| Ce/Ba | −0.13 ± 0.04 | −0.37 ± 0.04 | 0.006 | 0.23 | 32 |
| Pr/Ba | −0.17 ± 0.08 | −0.89 ± 0.07 | 0.03 | 0.17 | 26 |
| Nd/Ba | −0.07 ± 0.05 | −0.37 ± 0.04 | 0.17 | 0.06 | 36 |
| Sm/Ba | 0.00 ± 0.06 | −0.76 ± 0.06 | 0.99 | < 0.001 | 28 |
| Gd/Ba | 0.15 ± 0.05 | −0.75 ± 0.05 | 0.008 | 0.23 | 29 |
| Tb/Ba | 0.24 ± 0.13 | −1.64 ± 0.14 | 0.09 | 0.21 | 15 |
| Dy/Ba | 0.23 ± 0.06 | −0.71 ± 0.05 | < 0.001 | 0.36 | 30 |
| Ho/Ba | 0.12 ± 0.09 | −1.32 ± 0.10 | 0.22 | 0.08 | 21 |
| Er/Ba | 0.12 ± 0.05 | −0.81 ± 0.05 | 0.03 | 0.14 | 32 |
| Tm/Ba | 0.07 ± 0.07 | −1.61 ± 0.08 | 0.35 | 0.05 | 19 |
| Yb/Ba | 0.20 ± 0.07 | −0.97 ± 0.06 | 0.008 | 0.26 | 26 |
| Hf/Ba | 0.06 ± 0.07 | −1.06 ± 0.09 | 0.40 | 0.07 | 12 |
| Os/Ba | 0.26 ± 0.13 | −0.47 ± 0.16 | 0.07 | 0.29 | 12 |
| Pt/Ba | 0.20 ± 0.13 | −0.17 ± 0.15 | 0.17 | 0.29 | 8 |

**Table S2 The empirical baseline abundance pattern for lighter elements calculated from the abundances of the 13 stars with [Eu/Fe] ≤ +0.3.** We adopt a minimum uncertainty of 0.10 dex when the number of stars, $n$, used to compute the mean is ≤ 2. The columns list (1) the element symbol, (2) log of the abundance ratio, (3) standard error in this ratio, and (4) number of stars used to compute this ratio.

| Element X | log ε $(X/Zr)_{base}$ | standard error | n |
|---|---|---|---|
| Se | 0.64 | 0.10 | 2 |
| Sr | 0.16 | 0.04 | 13 |
| Y | -0.73 | 0.02 | 13 |
| Zr | 0.00 | 0.05 | 13 |
| Nb | -1.17 | 0.11 | 4 |
| Mo | -0.54 | 0.05 | 13 |
| Ru | -0.66 | 0.09 | 10 |
| Rh | -1.39 | 0.10 | 1 |
| Pd | -0.99 | 0.06 | 12 |
| Ag | -1.76 | 0.05 | 3 |
| Cd | -1.31 | 0.10 | 2 |
| Sn | -1.01 | 0.10 | 1 |
| Te | -0.12 | 0.10 | 1 |

**Table S3: Same as Table S2, but for heavier elements.**

| Element X | log ε $(X/Ba)_{base}$ | standard error | n |
|---|---|---|---|
| Ba | 0.00 | 0.05 | 13 |
| La | -0.87 | 0.06 | 11 |
| Ce | -0.32 | 0.05 | 8 |
| Pr | -0.80 | 0.11 | 7 |
| Nd | -0.33 | 0.05 | 12 |
| Sm | -0.70 | 0.08 | 5 |
| Eu | -1.15 | 0.03 | 13 |
| Gd | -0.77 | 0.08 | 6 |
| Tb | -1.58 | 0.10 | 1 |
| Dy | -0.70 | 0.11 | 7 |
| Ho | -1.27 | 0.19 | 4 |
| Er | -0.78 | 0.08 | 9 |
| Tm | -1.39 | 0.13 | 2 |
| Yb | -0.99 | 0.09 | 5 |
| Hf | -0.96 | 0.10 | 1 |
| Os | -0.35 | 0.10 | 1 |
| Pt | -0.31 | 0.10 | 1 |

**Caption for Data S1 (.csv file):**

**Stellar sample: model parameters, abundances, uncertainties, and literature references.**
The data are sorted by decreasing [Eu/Fe] ratios. See text for definitions. The columns list, in this order: the catalogue designation of each star; its right ascension and declination (J2000 equinox); its identifier in the Gaia DR3 catalogue (*31*); literature values of the effective temperature $T_{\text{eff}}$ (K), logarithm of surface gravity log $g$ (cgs units), microturbulent velocity parameter (km s$^{-1}$), model metallicity, and their uncertainties; our calculated $T_{\text{eff}}$ (K), log $g$ and their uncertainties; the ratios [Eu/Fe] and [Fe/H]; log ε values for the elements Se, Sr, Y, Zr, Nb, Mo, Ru, Rh, Pd, Ag, Cd, Sn, Te, Ba, La, Ce, Pr, Nd, Sm, Eu, Gd, Tb, Dy, Ho, Er, Tm, Yb, Hf, Os, Pt and Th; the uncertainty in [Fe/H]; the uncertainties in log ε of the elements Se to Th in the same order; reference for [Fe/H]; and references for log ε of the elements Se to Th in the same order.

Abbreviations used in the file for references:
Ao17 (*62*), Ba11 (*105*), Ca18 (*106*), Co02 (*107*), Co05 (*108*), De05 (*109*), Fr07 (*110*), Ha09 (*111*), Ha12 (*98*), Ha20 (*112*), Hi02 (*65*), Hi17 (*113*), Ho04 (*114*), Ho06 (*115*), Ho07 (*116*), Iv06 (*117*), J02a (*118*), J02b (*60*), La08 (*119*), Ma10 (*120*), Ni10 (*121*), Pe20 (*46*), Pl20 (*122*), R22a (*9*), R22b (*63*), RL12 (*123*), Ro10 (*124*), Ro12 (*125*), Ro14 (*126*), Ro18 (*38*), Sa18 (*127*), Si13 (*128*), Sn03 (*129*), Sn09 (*45*), Sp18 (*130*)

Note added by authors:
The arXiv preprint server does not allow the use of ancillary files when using a PDF upload. Table Data S1 can be obtained through the Science website at the link provided on the first page or by emailing the corresponding author.